\documentstyle[psfig,12pt]{article}
\newcommand{\eg}{{\sl e.g.}}

\newcommand{\etal}{{\sl et al.}}

\bf

\begin{document}
\noindent
{\Large\bf The large-scale distribution of warm ionized gas around
nearby radio galaxies with jet-cloud interactions}
\vglue 0.5cm\noindent
{\bf C.N. Tadhunter$^{1}$,  M. Villar-Martin$^{1,2}$, R. Morganti$^{3}$,
J. Bland-Hawthorn$^{4}$ \&
D. Axon$^{5}$}
\vglue 0.5cm\noindent
{\small $^{1}$ Department of Physics, University of Sheffield, Sheffield S3 7RH, UK \\
$^{2}$ Institude d'Astrophysique, 98bis Boulevard Arago, F-75014, Paris, France \\
$^{3}$ NFRA, PO Box 2, 7990 AA Dwingeloo, The Netherlands \\
$^{4}$ Anglo-Australian Observatory, PO Box, 296 Epping, NSW 2121, Australia \\
$^{5}$ Division of Physics and Astronomy, Department of Physical Sciences,
University of Hertfordshire, College Lane, Hatfield, Herts AL10 9AB, UK }

\vglue 0.5cm\noindent
{\bf Abstract} Deep, narrow-band H$\alpha$ observations taken with
the TAURUS Tunable Filter (TTF) on the 4.2m WHT telescope
are presented for two nearby radio
galaxies with strong  jet-cloud interactions.
Although the brightest emission
line components are closely aligned with the radio jets --- providing
nearby examples of the ``alignment effect'' most commonly observed
in high redshift ($z > 0.5$) radio galaxies --- lower surface brightness
emission line structures are detected at large distances (10's of kpc) from
the radio jet axis. These latter
structures cannot be reconciled with anisotropic illumination
of the ISM by obscured quasar-like sources, since parts of the  
structures lie outside any plausible quasar ionization cones. Rather, the
distribution of the emission lines around the fringes of the extended
radio lobes suggests that the gas is ionized either by direct interaction with
the radio components, or by the diffuse photoionizing radiation fields 
produced in the shocks generated in such interactions. These observations
serve to emphasise that the ionizing effects of the radio components can
extend far from the radio jet axes, and that deep emission line imaging
observations are required to reveal the true distribution of warm gas
in the host galaxies. We expect future deep imaging observations to 
reveal similar structures perpendicular to the radio axes in the high-z radio galaxies.

\section{Introduction}
Powerful radio galaxies are frequently associated with extended emission line
nebulosities which extend on radial scales of 5 --- 100 kpc from the nuclei
of the host early-type galaxies (Hansen \etal\ 1987, Baum \etal\ 1988).
The morphological and kinematical properties of these
nebulae provide important clues to the origins of
the gas, and the origins of the activity as a whole. The study of such gas is, for example, important for our understanding of the building of galaxy disks 
and bulges by infall since their epoch of formation.
Therefore, it is crucial to
determine the extent to which the observed emission line properties reflect
the intrinsic distribution of the warm gas in the haloes of the host galaxies,
and the extent to which they reflect the effects of the nuclear activity and
interactions with the extended radio sources.  
 
In most low redshift ($z < 0.2$) radio galaxies the optical emission line regions are broadly distributed in angle around the nuclei of
the host galaxies, the correlations between the optical
and radio structural axes are weak, and 
the gas kinematics are often quiescent, with line
widths and velocity shifts consistent in most cases with gravitational
motions in the host early-type galaxies (Tadhunter, Fosbury
\& Quinn 1989; Baum, Heckman \& van Breugel 1990). 
However, optical observations
reveal a dramatic change in the properties of the nebulosities as the
redshift and radio power increase: the emission line kinematics become 
more disturbed (compare Tadhunter \etal\ 1989 with McCarthy, Baum \&
Spinrad 1996) 
and the optical/UV structures become more closely aligned with the radio 
axes of the host galaxies (McCarthy \etal\ 1987,
McCarthy \& van Breugel 1989).  The most recent high resolution HST images 
of $z \sim 1$ radio galaxies show that
the structures are not only closely
aligned with the radio axes, but they are also highly collimated, with
a jet-like
appearance (Best et  al. 1996). 
The nature of  the ``alignment effect'' is
a key issue for our general understanding of powerful radio galaxies, of particular relevance to the use of radio sources as probes of the high redshift universe.

Of the many models which have been proposed to explain the alignment effect, the
two which have received the most attention are the {\it anisotropic illumination}
and the {\it jet-cloud interaction}\footnote{We use ``jet-cloud interaction''
as a generic term to describe interactions between the warm ISM
and the radio-emitting components, which could include the radio lobes
and hot-spots, as well as the radio jets.} models. 

In the case of anisotropic illumination it is proposed that
the gaseous haloes of the host galaxies are illuminated by the broad
radiation
cones of the quasars hidden in the cores of the galaxies (\eg\ Barthel
1989), with the emission lines resulting from photoionization of the ambient ISM by the EUV radiation in the cones (\eg\ Fosbury 1989), and the extended optical/UV
continuum comprising a combination of the nebular continuum emitted by the warm
emission line clouds (Dickson \etal\ 1995) and scattered quasar light 
(Tadhunter \etal\ 1988, Fabian 1989). The alignment of the obscuring tori perpendicular to the collimation axes of the plasma jets then leads to a 
natural alignment of the extended nebulosities with the radio axes. The best
evidence to support the anisotropic illumination model is provided by
polarimetric observations of powerful radio galaxies at all redshifts
which show evidence for high UV polarization and scattered quasar features
(\eg\ Tadhunter \etal\, 1992, Young \etal\ 1996, Dey \& Spinrad 1996, Cimatti \etal\ 1996, Cimatti
\etal\ 1997, Ogle \etal\ 1997).
It would be difficult to explain these polarimetric results in terms of
any mechanism other than scattering of the anisotropic radiation
field of an illuminating quasar or AGN. However, despite the success
of the illumination model at explaining the polarization properties, a 
sigificant fraction of radio galaxies --- comprising $\sim$30 -- 50\% of radio
galaxies at $z\sim1$, and a smaller proportion of lower redshifts --- are dominated by jet-like UV emission line structures,
which are  more
highly collimated than would be expected on the basis of the
45 -- 60$^{\circ}$ opening half-angle illumination cones
predicted by the unified schemes for powerful radio galaxies
(Barthel 1989, Lawrence 1991). Moreover,
the highly disturbed emission line kinematics
observed in many high-z sources are also difficult to reconcile with
quasar illumination of the undisturbed ambient ISM of the host galaxies.

Jet-cloud interactions  
have the potential to explain many of the
features of powerful radio galaxies which cannot be explained in terms
anisotropic quasar illumination. Although the jet-cloud interactions
are likely to be complex, at the very least the clouds will be compressed,
ionized and accelerated as they enter the shocks driven through
the ISM by the jets. Therefore, jet-cloud interactions provide a promising
explanation for the high-surface-brightness and extreme emission line kinematics
of the structures aligned along the radio axes of high-z sources.  
Indeed, recent spectroscopic observations of jet-cloud interactions in 
low redshift
radio galaxies provide clear observational evidence for the acceleration and ionization of warm clouds by the jet-induced shocks (\eg\ Clark \etal\ 1998, Villar-Martin \etal\ 1999). Moreover, theoretical modelling work
has demonstrated that jet-induced shocks are a viable,
if not unique,
mechanism for producing the emission line spectra of radio galaxies
(\eg\ Dopita \& Sutherland 1998).  

It is clear that no single mechanism can explain the emission line
properties of radio galaxies over the full range of redshift and radio
power; some combination of AGN illumination and jet-cloud interactions is required, with the jet-cloud interactions becoming increasingly important
as the redshift and/or radio power increases. However, a major problem with
such a combined model is that, 
while the polarimetric results demonstrate that quasar illumination is 
important in many high redshift sources, the extended structures are often
dominated by highly collimated jet-like structures, with no
sign of the  broad cone-like emission line distributions predicted
by the unified schemes for powerful radio galaxies. 

How do we explain this
dearth of broad cones in the objects with the
most highly collimated structures? Possibilities include the following.
\begin{itemize}
\item {\bf The gas structures are intrinsically aligned along the radio
axes of the high redshift sources}, so that the emission 
line nebulae reflect the true 
distribution of warm/cool gas, rather than the ionization patterns
induced by the jets or illuminating AGN. 
For example, West (1994)
has proposed that a general alignment of the gas structures along the
radio axis may be a natural consequence of the formation of giant elliptical
galaxies in a heirarchical galaxy formation scenario, although it is
not clear that the structures formed in this way would be quite as
highly collimated as those observed in the high-z radio galaxies. 
\item {\bf The nuclei of the host galaxies do not contain powerful
quasars}, and the ionization of the extended gas is dominated by the jets: either by direct jet/cloud interactions, or by illumination by the
relativistically-beamed jet radiation. This scenario is supported
by the discovery at low redshifts of a class of powerful radio galaxies with weak, low ionization nuclear
emission line regions (see Laing 1994, Tadhunter \etal\ 1998). 
However, at least some high-z radio galaxies with highly
collimated UV structures show direct evidence for powerful quasar
nuclei in the form of scattered quasar features in their polarized
spectra, so this explanation cannot hold in every case.   
\item {\bf The broad-cone radiation of the buried quasars does
not escape from the nuclear regions of the host galaxies}, because
of the absorbing effects of circum-nuclear gas. In this case the ionization of the
extended gas in the aligned structures is likely to be dominated by jet-cloud interactions, but quasar or beamed jet radiation may also contribute along
the jet axis,
if the jets punch a hole in the obscuring material. Direct evidence for
requisite obscuring material in the central regions is provided by the relatively high occurrence rate of 
associated CIV and Ly$\alpha$ absorption line systems in the UV
spectra of radio-loud
quasars  
(\eg\ Anderson \etal\ 1987, Wills \etal\ 1995), the relatively
red SEDs of steep spectrum radio quasars (Baker 1997), and the detection of
significant BLR reddening in a substantial fraction of nearby
broad-line radio galaxies (\eg\ Osterbrock, Koski \& Miller 1995, Hill
\etal\, 1996).
\item {\bf The dearth of broad ionization cones in the high-z sources
is a consequence of an observational selection effect:} most of the 
existing emission line
images of the high-z sources have been taken in the light of the
low ionization [OII]$\lambda$3727 line which is emitted particularly strongly
by the jet-induced shocks (\eg\ Clark \etal\ 1997, 1998), 
but is relatively weak in  the more highly ionized
quasar illumination cones. Thus, given that the 
published ground-based images of the high-z objects are relatively
shallow and have a low spatial resolution, while the published HST images have a higher
resolution but are insensitive to low surface brightness structures,
the existing images are likely to be biased in
favour of the high-surface-brightness shocked structures along the radio axes. In this
case, we would expect deep emission images to reveal gaseous structures outside
the main high surface brightness structures aligned along the radio axes.    
If the gas away from the radio axis is predominantly photoionized by quasars
hidden in the cores of the galaxies we would expect the extended low surface
brightness structures to have a broad distribution, 
consistent with quasar illumination. Detection of such
emission line morphologies in the objects with the most highly
collimated structures would lead to a reconciliation between the anisotropic
illumination and jet-cloud interaction models, thereby resolving the outstanding uncertainties surrounding the nature of the alignment effect.
\end{itemize}
In order to test the latter possibility
it is important to obtain deep emission line imaging observations for the objects with closely aligned radio and optical structures.
We report here pilot observations of two intermediate-redshift radio galaxies
--- 3C171 ($z=0.2381$) and 3C277.3($z=0.08579$) --- which are nearby prototypes
of the high-z radio galaxies, in the sense that they show high surface 
brightness emission line structures which are closely aligned along their radio
axes. The results challenge some commonly-held assumptions about the
ionization of the extended gaseous haloes around powerful radio galaxies.

\section{Observations}

Emission line and continuum observations of
3C171 and 3C277.3(Coma A)
were taken using the Taurus Tunable Filter (TTF) on the
4.2m WHT telescope at the La Palma Observatory on the night of the 
27th January 1998. A log of the observations is presented in 
Table 1, while a
full description of the TTF is given in Bland-Hawthorn \& Jones (1998a,b). Use of the
f/2 camera of TAURUS with the Tek5 CCD resulted in a pixel scale of 0.56
arcseconds per pixel; and
the seeing conditions were subarcsecond for the observations reported here.
The faintest structures visible in the images for both objects have
an H$\alpha$ surface brightness of $\sim1\times10^{-17}$ erg cm$^{-2}$ 
s$^{-1}$ arcsec$^{-2}$.

Because of ghosting effects in the flat field images, no flat fielding of the
data was attempted. However, comparisons between images taken with different
filters and/or with the objects placed in different positions on the detector,
demonstrate that the ghost images of stars and galaxies in the field do not contaminate the images of the main
target objects described below.
 
The reduction of the images consisted of 
bias subtraction, atmospheric extinction correction, flux calibration,
sky subtraction and registration. From the comparison 
between the measurements of the flux calibration standard stars
taken at various times during the run it is estimated
that the absolute flux calibration is accurate to within $\pm$30\%, and
the H$\alpha$ emission line fluxes agree at this level with the long-slit
spectroscopy measurements  in Clark (1996). For the
emission line images, the TTF was tuned to the wavelength of H$\alpha$ shifted to the redshift of the emission lines in the nuclear regions of the galaxies.
However, velocity structure in the haloes of the host galaxies may result in
the emission lines in the extended structures not being exactly centred in the
TTF bandpass, which has a Lorentzian shape. We estimate that, at maximum, this
will result in the fluxes being underestimated by a factor of two for components
with extreme $\pm$600 km/s shifts, but this will not affect our main conclusions
which are based largely on the emission line structures, rather than the
emission line fluxes.

In order the facilitate comparison between the radio and optical structures, radio images were obtained for both sources. The radio and optical
images were registered by matching the positions of the core radio sources 
with the positions of the continuum
centroids in the optical continuum images, with the pixel scale and rotation
of the optical images calibrated using the known positions of stars in the CCD
fields.

The radio image of Coma A was made using data taken with the VLA  A-array
configuration at 1.4 GHz (20cm). This gives a resolution for the final image 
of 1.14x1.13 arcseconds in p.a. -72.
The data, which were extracted from the VLA archive, were originally 
presented and discussed in great detail by van Breugel \etal\ (1985). We
therefore refer to that paper for all the radio information about Coma A.

The radio image of 3C171 was kindly provided by K. Blundell. The image 
was made with the VLA at 8~GHz with a resolution of 1.3 arcsec FWHM.
More information about the radio characteristics of this source can be found in Blundell
(1996).

\section{Results}
\subsection{3C277.3 (Coma A)}

Previous spectroscopic and imaging observations of 3C277.3 by
van Breugel \etal\ (1985) and Clark (1996) show the presence
of a series of high surface brightness structures along the
radio axis. These include: a high ionization emission line region
associated with knots in radio jet some 6 arcseconds to the 
south east of the nucleus; an enhancement in the emission line 
flux close to the hotspot in the northern radio lobe; and an emission
line arc which partially circumscribes the northern radio hotspot.
Although the kinematic and ionization evidence for a jet-cloud
interaction in this source is less clear than in some other radio
galaxies (\eg\ 3C171: see below) --- the ionization state is relatively
high and the emission lines relatively narrow --- van Breugel
\etal\ (1985) found evidence for a jump in the emission line radial velocities
across the northern radio lobe, while Clark (1996) noted that the ionization
has a minimum, and the electron temperature a peak, at the position of the
northern radio hotspot. Note that there is no clear evidence for a powerful
quasar nucleus in this source: the nuclear regions show no evidence
for scattered quasar light, and the nuclear emission line region has a 
relatively low luminosity and ionization state compared with the brightest
extended emission line regions along the radio axis.

Our deep H$\alpha$ images (Figure 1a) show that the emission line regions
along the radio axis
form part of a spectacular system of interlocking emission line arcs and filaments, which
extend almost as far perpendicular as parallel
to the radio axis. Of particular interest is the fact that the brightest
arc structure wraps a full 180$^{\circ}$ around the nucleus, with enhancements
in the emission line surface brightness where the arc intercepts the radio
axis to the north and south of the nucleus. The spatially integrated
H$\alpha$
fluxes of the bright knots along the radio axis (including the nucleus),
the extended low surface brightness filaments, and the nebula as a whole
are $2.5\times10^{-14}$, $1.6\times10^{-14}$ and $4.1\times10^{-14}$ erg 
s$^{-1}$ cm$^{-2}$ respectively. For our adopted cosmology\footnote{$H_0 = 50 km s^{-1} kpc^{-1}$ and $q_0 = 0.0$ assumed throughout.} 
the corresponding H$\alpha$ emission line
luminosities are $8.6\times10^{41}$, $5.6\times10^{41}$ and $1.42\times10^{42}$
erg s$^{-1}$ respectively. 

The fact that the main arc and filament structures are not visible, or
are considerably fainter, in the
intermediate-band
continuum image (Figure 1b) --- which is at least as sensitive to continuum structures
as the narrow-band H$\alpha$ image --- demonstrates that they are predominantly
emission line structures. However, 
a number of faint galaxies and continuum structures are detected within
100 kpc of the nucleus of Coma A, and at least some of these continuum structures (highlighted by arrows in the figure) are intimately associated with the
extended H$\alpha$ filamentary structures.

Overall, the Coma A system has the appearance of an interacting group of
galaxies: the H$\alpha$ filaments bear a marked resemblance to 
the HI tails detected in 21cm radio observations
of interacting groups (\eg\ the M81 group: Yun \etal\ 1994); and it is
plausible that the faint continuum
structures represent the debris of interactions/mergers between the
dominant giant elliptical galaxy and less massive galaxies in the same group.
The X-ray luminosity of Coma A ($L_{0.5-3kev} < 8.1\times10^{42}$ erg s$^{-1}$:
Fabbiano \etal\ 1984) is also consistent with a group environment.   

Figure 2 shows an overlay of the emission line image and the  6cm radio map. 
This reveals a striking match between the emission line
and radio structures. As well as the high surface brightness features along
the radio axis, the brightest arc to the north of the nucleus closely
follows the outer edge of northern radio lobe. The emission line 
structures appear to bound the radio structures: the brighter
emission line features have a similar radial and lateral extent to the radio
features. It is notable, however, that fainter, more diffuse emission line structures
are detected well outside the radio lobes on the
northern and eastern sides of the galaxy.

The detection of arc structures circumscribing radio lobes is not
without precedent: the intermediate redshift radio galaxies PKS2250-41 (Clark \etal\ 1998,
Villar-Martin \etal\ 1999), 3C435A (Rocca-Volmerange \etal\ 1994) and PKS1932-46
(Villar-Martin \etal\ 1998), the high redshift radio galaxies 3C280
(McCarthy \etal\ 1995) and 3C368 (Best \etal\ 1996), and the central
elliptical galaxy in the cooling flow cluster A2597 (Keokemoer \etal\ 1999),
all show evidence for arcs associated with radio lobes. In many of these
cases there is also spectroscopic evidence that the emission line gas extends beyond the radio lobes.


\subsection{3C171}

3C171 is another example of an object in which high surface brightness emission
line structures are closely aligned along the axis of the radio jets
(Heckman \etal\ 1984, Baum \etal\ 1988). The 
spectroscopic evidence for a jet-cloud interaction in this source is strong:
the emission line kinematics along the radio axis are highly disturbed; and the the general
line ratios and ionization minima coincident with the radio hotspots to the
east and west of the nucleus provide strong evidence that the emission line
gas has been compressed and ionized by  jet-induced shocks
(Clark \etal\ 1998). A further
possible consequence of the jet-cloud interactions is the highly disturbed 
radio structure, with the radio lobes showing a greater
extent perpendicular- than parallel to the jet axis, giving
an overall H-shaped appearance (Heckman \etal\ 1984, Blundell 1996). 

Our deep H$\alpha$ and continuum images of this source are shown in Figure 3, while an overlay of the optical emission line image and
the radio map is presented in Figure 4. From the continuum-subtracted H$\alpha$
image we measure spatially integrated emission line fluxes of
$2.12\times10^{-14}$, $6.2\times10^{-16}$ and $2.63\times10^{-14}$  
erg s$^{-1}$ cm$^{-2}$ 
respectively for
the high surface brightness structures aligned along the radio axis
(including the nucleus), the
faint filament to the north, and the nebula as a whole. The corresponding
H$\alpha$ emission line luminosities are $6.6\times10^{42}$, 
$1.9\times10^{41}$ and  $8.1\times10^{42}$  erg s$^{-1}$ respectively.

Although the emission line structures in 3C171 are
clearly different in detail from those detected in Coma A, there are important
general similarities. Most notably, as in Coma A, the highest surface brightness emission line
features are closely aligned along the radio axis, yet lower surface brightness
structures are also detected in the direction perpendicular to the radio axis. The emission line structures have a similar radial extent in the
directions perpendicular and parallel to the radio axis. Away from the
radio axis, the most striking emission line feature is the filament 
which extends 9 arcseconds (45 kpc)
north of the nucleus. This feature lies along the
fringes of the western radio lobe, just as the arc to the north of the nucleus
in Coma A skirts the outer edge of the northern radio lobe in that object.
A further similarity with Coma A is that, in the radio axis direction, the
radio structures are confined within the emission line structures, which have
a similar radial extent. We also find evidence
for emission line gas that is not clearly associated with radio structures: 
the faint, diffuse
H$\alpha$ emission to the south 
east of the nucleus lies well to the south of the extended eastern radio
lobe. However, 3C171 is different from Coma A in the sense that
the radio lobes extend further than the emission line structures in the
direction perpendicular to the radio axis on the northern side of the galaxy.

\section{Discussion}

The main aim of the deep emission line imaging
observations
was to attempt to detect the broad emission line cones outside the main
aligned structures, and thereby reconcile the AGN illumination and jet-cloud
interactions models. The unified schemes predict illumination 
cones with opening 
half-angles
of 45-60$^{\circ}$. Although the extended emission line nebulosities in low 
redshift radio galaxies
rarely show the  sharp-edged cone structures observed in some Seyfert galaxies
(\eg\ Pogge 1988, Tadhunter \& Tsvetanov 1989), the emission 
line distributions are generally consistent with broad cone illumination
of an inhomogeneous ISM 
(Hansen \etal\ 1987, Baum \etal\ 1989, Fosbury 1989). The detection of similar
emission line distributions in 3C171 and
Coma A would support the idea that the extended ionized haloes are photoionized
by quasars hidden in the cores of the galaxies.

The deep imaging observations presented in this paper have confounded our
expectations in the sense that, while they do show extended 
of emission line gas well away from the radio axis, the emission line
distribution cannot be reconciled with any plausible ionization cone model.
Not only do some of the features wrap through a full 180$^{\circ}$ in position angle around the nucleus of Coma A, but there are
no sharp boundaries in the surface brightness of the structures, corresponding
to the edges of an ionization cone. It is possible for the emission line 
distributions to appear broader than the nominal  45-60$^{\circ}$ cones
predicted by the unified schemes if the cone axes are tilted towards
the observer. However, in order to explain the emission line distributions
in 3C171 and Coma A in this way, the cones would have to be tilted to such
an extent that the observer's line of sight would lie within the cone
and we would see the illuminating AGN directly. Clearly this is not the case,
and it appears highly unlikely that the extended filaments away from
the radio axis are
photoionized by a central source of ionizing photons.  

The most plausible alternative to quasar illumination is ionization by the
shocks associated with the expanding radio jets and lobes. The emission lines
could be produced as the warm clouds cool behind the shock fronts or, alternatively, as a consequence of photoionization of precursor clouds by
the ionizing photons produced in the cooling, shocked gas. In either case we would 
expect
a close morphological association between the radio and optical 
structures, just
as we observe in 3C171 and Coma A. By adapting equation 4.4 of Dopita and Sutherland (1996), and assuming a shock speed through the warm clouds of
200 km s$^{-1}$, we estimate that the rate of flow of warm ISM through the
shocks would have to be at least 1.9$\times$10$^4$ M$_{\odot}$ yr$^{-1}$
for 3C171, and 3.2$\times$10$^3$ M$_{\odot}$ yr$^{-1}$ for Coma A, in
order for the emission line luminosities of the
nebulae as a whole to be produced entirely by shock ionization. Energetically,
the shock ionization mechanism appears to be feasible in the sense that
the total emission line luminosities of the sources
are $<$10\% of
the bulk powers of the radio jets (Clark 1996, Clark \etal\ 1998)\footnote{In order to derive this result we have scaled the
results of Clark (1996), who considered only the emission line components
along the radio axis, to the total emission line fluxes for the nebulae
as a whole, as derived from our H$\alpha$ images.}.


However, it is not possible to rule out some
contribution to the ionization of the extended structures
by a central photoionizing source. As discussed in the
introduction, some radio sources with relativistic jets
may not have powerful quasar nuclei. If this is the case, the narrow
beams of radiation emitted by the jets could contribute to the ionization of the structures along
the radio axis,
although the ionization of the more extended filamentary structures would
continue to be  be dominated by interactions with the rdaio lobes.

One further possibility is that the structures are photoionized by young
stars associated with the filaments. This is supported by
the presence of faint continuum structures associated with the H$\alpha$
filaments (see Figure 1(c)), and the spectroscopic detection of excess UV continuum emission to the
north and south of the nucleus along the radio axis above the level expected
for the nebular continuum emitted by the warm gas
(Clark 1996). Without further information
on the nature and spectrum of the extended continuum structures it is difficult
to test this model at the present time.

An open question for both 3C171 and Coma A is the extent to which the structures
reflect the true distributions of ionized gas in the haloes of the host galaxies,
and the extent to which the structures are distorted by their interaction
with the radio components. It is possible for shock fronts to sweep up material
into shell-like structures. However, given that the clouds are likely to be destroyed by hydrodynamical interactions
with the fast, hot wind behind the shock fronts within
a few shock crossing times (\eg\ Klein, McKeee \& Collella 1994), 
and given also the presence of
diffuse H$\alpha$ emission well away from the radio structures in
both Coma A and 3C171, 
it seems more plausible that these 
represent pre-existing gas structures. Cloud destruction by
shocks may also lead to a relative absence of warm gas in 
the lobes, further enhancing the shell-like appearance of the emission
line structures.
In the 
case of Coma A it is likely that we are seeing
the results of interactions between between the radio-emitting
components and the gaseous remnants of mergers/interactions in a group
of galaxies. 

Clearly, detailed measurements of the kinematics, 
line ratios, and continuum spectra of
the filamentary structures are required in order to resolve
the outstanding issues concerning the physical state, ionization
and origins of the warm gas.

\section{Implications for high redshift radio galaxies}

Our observations demonstrate the presence of extended gaseous structures
well away from the high-surface-brightness structures aligned along
the radio axes in two nearby radio
galaxies. Given that Coma A and 3C171 are similar to the high redshift
radio galaxies in the sense that they show high-surface-brightness emission line
structures closely aligned along their radio axes, as well as evidence
for disturbed emission line kinematics, it seems likely that similar extended 
gaseous structures also exist in the high-z sources. In this case, the highly
collimated
structures visible in the existing images of some $z \sim 1$ 
3C radio sources may reflect more the ionization pattern induced by the radio jets than the
true distribution of warm/cool gas in the host galaxies. 

Note, however,  that 3C171 and Coma A
have radio and emission line luminosities that are an order of magnitude
lower than 3C radio galaxies at $z \sim 1$. Furthermore, the radio lobes in
3C171 and Coma A extend further in the direction perpendicular to the radio
jets than is typical in high redshift 3C radio sources.  
Therefore, it is difficult to predict the detectability of the extended low surface brightness
structures in the high-z radio galaxies ($z > 1$) based on a straightforward extrapolation of the properties of 3C171 and Coma A. 
Given the smaller lateral extents of the radio lobes in the 
high-z sources, the ionization effects associated with the lobes may be less
effective at large distances from the radio axes in such objects. In
addition, the structures in the high-z sources will
be 
subject to $(1+z)^{-4}$ cosmological surface brightness dimming which
will make them more difficult to detect relative to nearby sources for
similar intrinsic brightness levels. However, set against this is the
fact that, in contrast to 
Coma A and 3C171, there exists good polarimetric evidence that many 
of the high-z
radio galaxies contain powerful quasars hidden in their cores. Provided
that the ionizing photons in the broad ionization cones can
escape the nuclear regions (but see discussion in introduction),
illumination by the quasar cones will enhance the surface
brightnesses of the extended structures and render them more easily
detectable.

The extended low surface brightness structures may already have been
detected spectroscopically in at least one high-z source: deep, long-slit
Keck spectra taken along the radio axis
of 3C368 ($z = 1.135$) by Stockton, Ridgway \& Kellogg (1996) show the presence of a faint  emission 
line region well
outside the main high surface brightness  emission line
regions closer to the nucleus. The  relatively
narrow lines and high ionization state measured in this
faint, low-surface-brightness region are consistent
with quasar illumination of the undisturbed ambient medium of the host galaxy.

Some encouragement may also be drawn from the detection of large Ly$\alpha$
haloes around radio galaxies at $z > 2$ (\eg\ Adam et al. 1997). Although the Ly$\alpha$ in these
haloes may  not be formed by direct photoionization by an AGN, but rather
by resonant scattering of Ly alpha photons produced in the 
extended regions around
the nuclei (Villar-Martin \etal\  1996), these observations at least demonstrate the presence of 
extensive haloes of cool ISM surrounding the host galaxies of some of the highest
redshift radio galaxies.       

Thus, we
expect future deep emission line imaging of $z \sim 1$ radio galaxies to reveal
the true distribution of the extended ionized gas in the host galaxies, and to provide clues to the origins of the gas and the evolution of the host galaxies.

\section{Conclusions}

Deep emission line imaging observations of two nearby examples of 
the radio-optical alignment effect have revealed extensive low-surface-brightness emission line structures well away from the radio
axes, thus demonstrating that the intrinsic distribution of warm gas is more
extensive than previosly suspected. 

The general distribution of the gaseous structures is imcompatible with
the standard quasar illumination picture, while their association with
the extended radio structures provides clear evidence that they are interacting
with the radio lobes, hotspots and jets. These may be objects in which
the ionization of the extended emission line regions is entirely dominated by shocks
induced by interactions between the radio plasma and the ISM.

It is often assumed that broad distribution of ionized gas observed
in low redshift radio galaxies without clear signs of jet-cloud
interactions imply illumination by the broad ionization
cones of quasars hidden in the cores of the galaxies. These new observations
suggest that this may not always be the case, and that the lobes as well as
the jets may have a significant ionizing effect.
\vglue 0.5cm\noindent
{\bf Acknowledgments.} The Willian Herschel Telescope is operated on
the island of La Palma by the Isaac Newton Group in the Spanish
Observatorio del Roches de los Muchachos of the Instituto de Astrofisica
de Canarias. We thank Katherine Blundell for allowing us
to use to her radio image of 3C171. MVM acknowledges support from PPARC. 
\vglue 1.0cm\noindent
{\bf References}
\begin{description}


\item Adam, G., Rocca-Volmerange, B., Gerard, S.,
Ferruit, P., Bacon, R., 1997, A\&A, 326, 501

\item Anderson, S.F., Weymann, R.J., Foltz, C.B., Chaffee, F.H., 1987, AJ, 94, 278

\item Baker, J.C., 1997, MNRAS, 286, 23

\item Barthel, P.D., 1989, ApJ, 336, 606

\item Baum, S.A., Heckman, T.M., Bridle, A.H., van Breugel, W., Miley, G.K.,
1988, ApJS, 68, 833

\item Baum, S., Heckman, T., 1989, ApJ, 336, 702

\item Baum, S.A., Heckman, T.M., van Breugel, W., 1990, ApJS, 74, 389

\item Best, P.N., Longair, M.S., Rottgering, H.J.A., 1996, MNRAS, 280, 
L9.


\item Bland-Hawthorn, J. \& Jones, D.H. 1998a, PASA, 15, 44

\item Bland-Hawthorn, J. \& Jones, D.H. 1998b, In Optical Astronomical Instrumentation, SPIE vol. 3355, 855

\item Blundell K.M. 1996, MNRAS 283, 538




\item Cimatti, A., Dey, A, van Breugel, W., Antonucci, R., Spinrad, H., 1996, ApJ, 465, 145

\item Cimatti, A., Dey, A., van Breugel, W., Hurt, T., Antonucci, R., 1997, ApJ, 476, 677

\item Clark, N.E., 1996, PhD thesis, University of Sheffield.

\item Clark, N.E., Tadhunter, C.N., Morganti, R., Killeen, N.E.B., Fosbury, R.A.E., 
Hook, R.N., Shaw, M., 1997, MNRAS, 286, 558

\item Clark, N.E., Axon, D.J., Tadhunter, C.N., Robinson, A., O'Brien, P., 1998, ApJ, 494, 546

\item Dey, A., Spinrad, H., 1996, ApJ, 459, 133



\item Dickson, R.D., Tadhunter, C.N., Shaw, M.A., Clark, N.E.,
Morganti, R., 1995, MNRAS, 273, L29

\item Dopita, M.A., Sutherland, R.S., 1996, ApJS, 102, 161


\item Fabian, A.C., 1989, MNRAS, 238, 41p

\item Fabbiano, G., Miller, L., Trinchieri, G., Longair, M., Elvis, M., 1984,
ApJ, 277, 115

\item Fosbury, R.A.E., 1989: In: {\em ESO Workshop on Extranuclear Activity in Galaxies}, Meurs \& Fosbury (eds), p169

\item Hansen, L., Norgaard-Nielsen, H.U., Jorgensen, H.E., 1987, A\&ASuppl.,
71, 465

\item Heckman, T.M., van Breugel, W.J.M., Miley, G.K., 1984, ApJ, 286, 509




\item Hill, G.J., Goodrich, R.W., DePoy, D.L., 1996, ApJ, 462, 162

\item Klein, R., McKee, C., Colella, P., 1994, ApJ, 420, 213

\item Knopp, G.P., Chambers, K.C., 1997, ApJS, 109, 367

\item Koekemoer, A.M., O'Dea, C.P., Sarazin, C.L., McNamara, B.R., Donahue, M.,
Voit, G.M., Baum, S.A., Gallimore, J.F., 1999, AJ, in press

\item Lawrence, A., 1991, MNRAS, 252, 586

\item Longair, M.S., Best, P.N., Rottgering, H.J.A., 1995, MNRAS, 275, L47

\item McCarthy, P.J., van Breugel, W., Spinrad, H., Djorgovski, S., 1987, ApJ, 321, L29

\item McCarthy, P.J., van Breugel, W., 1989, in {\it The Epoch of Galaxy Formation}, ed. C. Frenk, Kluwer Academic Press, p57

\item McCarthy, P.J., Spinrad, H., van Breugel, W.J.M., 1995, ApJSupp., 99, 27

\item McCarthy, P.J., Baum, S., Spinrad, H., 1996, ApJS, 106, 281


\item Ogle, P.M., Cohen, M.H., Miller, J.S., Tran, H.S.,
Fosbury, R.A.E., Goodrich, R.W., 1997, ApJ, 482, 370

\item Osterbock, D.E., Koski, A.T., Phillips, M.M., 1976, ApJ, 206, 898

\item Pogge, R.W., 1988, 328, 519





\item Rocca-Volmerange, B., Adam, G., Ferruit, P., Bacon, R., 1994,
A\&A, 292, 20

\item Stockton, A., Ridgway, S.E., Kellogg, M., 1996, AJ, 112, 902



\item Tadhunter, C.N., Fosbury, R.A.E., di Serego Alighieri, S., 1988, in
Maraschi, L., Maccacaro, T., Ulrich, M.H., eds, Proc. Como Conf. 1988, BL Lac
Objects, Springer-Verlag, Berlin, p.79

\item Tadhunter, C.N., Fosbury, R.A.E., Quinn, P., 1989, MNRAS, 240, 255

\item Tadhunter, C., Scarrott, S., Draper, P., Rolph, C., 1992, MNRAS, 256, 53p

\item Tadhunter, C.N., Tsvetanov, Z., 1989, Nat, 341, 422

\item Tadhunter, C.N., Morganti, R., Robinson, A., Dickson, R., Villar-Martin,
M., Fosbury, R.A.E., 1997, MNRAS, submitted.

\item Villar-Martin, Binette, Fosbury, 1996, A\&A, 312, 751

\item Villar-Martin, M., Tadhunter, C.N., Morganti, R., Clark, N., Killeen, N.,
Axon, D., 1998, A\&A, 332, 479

\item Villar-Martin, M., Tadhunter, C.N., Morganti, R., Axon, D., 1999,
MNRAS, 307, 24

\item van Breugel, W., Miley, G., Heckman, T., Butcher, H., Bridle, A., 1985,
ApJ, 290, 496

\item West, M.J., 1994, MNRAS, 268, 79

\item Wills, B.J., Thompson, K.L., Han, M., Netzer, H., Wills, D., Baldwin, J.A.,
Ferland, G.J., Browne, I.W.A., Brotherton, M.S., ApJ, 447, 139

\item Young, S., Hough, J.H., Efstathiou, A., Wills, B.J., Axon, D.J., Bailey,
J.A., Ward, M.J., 1996, MNRAS, 279, L72

\item Yun, M.S., Ho, P.T.P., Lp, K.Y., 1994, Nat, 272, 530
\end{description}

\clearpage
\begin{table}
\begin{center}
\begin{tabular}{lccccl}
Object &Blocking Filter(\AA) &Etalon(\AA) &Exposure Time(s) &Seeing('') &Comments \\
 &$\lambda_c$/$\Delta \lambda$  &$\lambda_c$/$\Delta \lambda$  
& &FWHM  & \\ \\
3C171 &8140/330 &8124/28 &1800 &0.95 &H$\alpha +$Continuum \\
&8570/400 &--- &900 &0.95 &Continuum \\
3C277.3 &7070/260 &7124/19 &2$\times$900 &0.98 &H$\alpha +$Continuum \\
&7580/280 &--- &900 &0.98 &Continuum \\
\end{tabular}
\caption{Details of the TTF imaging observations for 3C171
and 3C277.3. The second and third columns give the central wavelengths/bandwidths for the blocking filters and the etalon respectively.}
\end{center}
\end{table}
\newpage\noindent
\begin{figure}
\psfig{figure=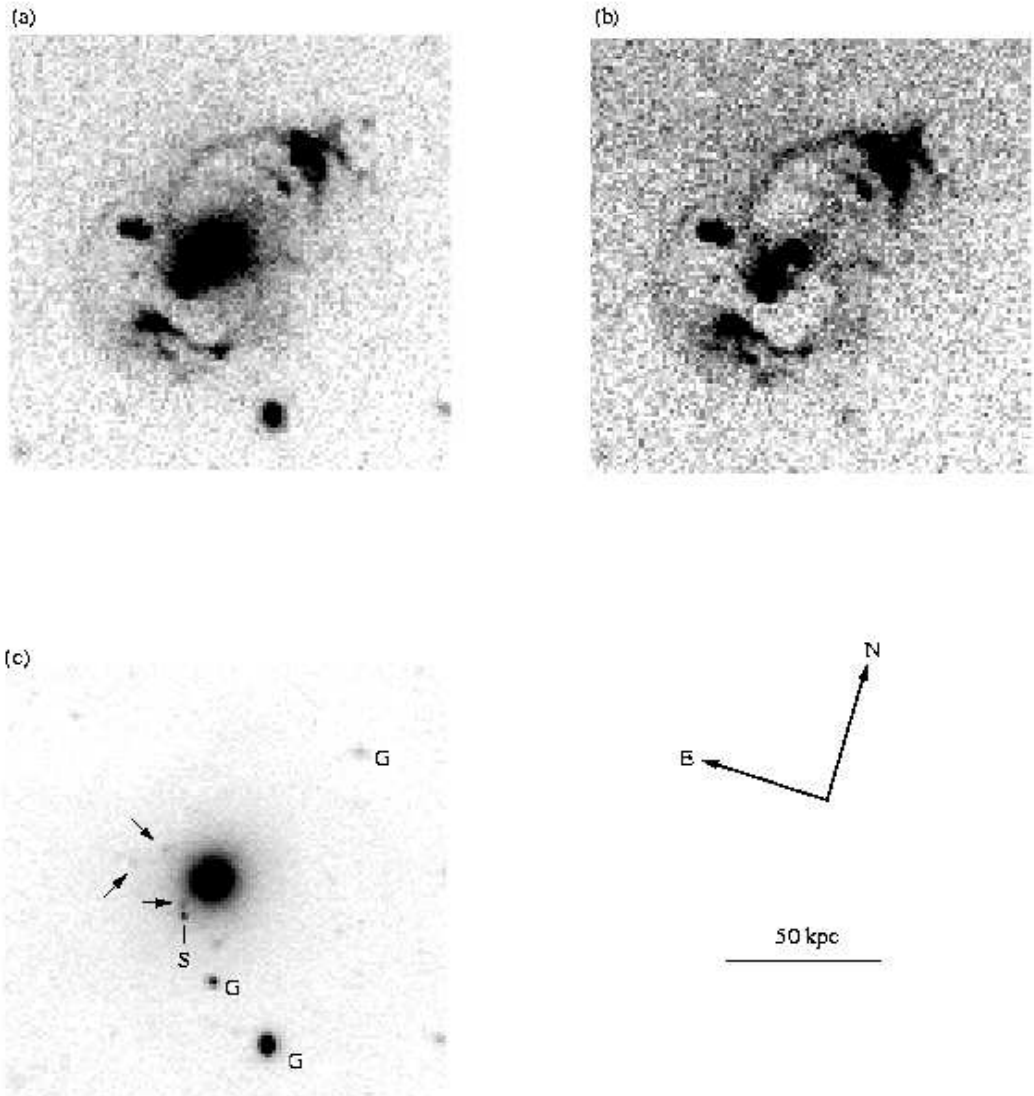,width=15cm}
\caption{TTF images of Coma A (3C277.3): (a) H$\alpha+$continuum; (b) pure  H$\alpha$; (c) continuum ($\lambda_c =$7580\AA\,). In the continuum image
the arrows point to faint continuum features associated with the extended
H$\alpha$ filamentary structures, the ``G'' symbols indicate galaxies, while
the ``S'' symbol indicates a faint star (unresolved in HST images).
}
\end{figure}
\begin{figure}
\psfig{figure=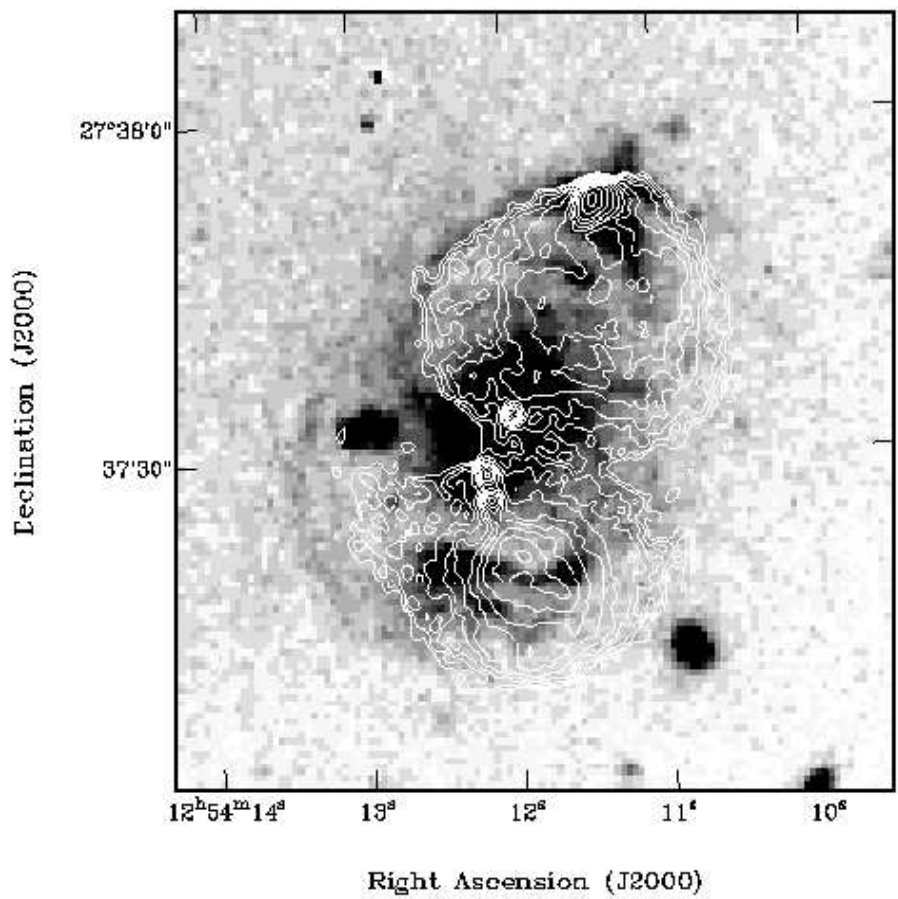,width=15cm}
\caption{Overlay of the H$\alpha+$continuum image for Coma A (greyscale) 
with the
6cm radio map of van Breugel \etal\, (1985) (contours).
}
\end{figure}

\begin{figure}
\psfig{figure=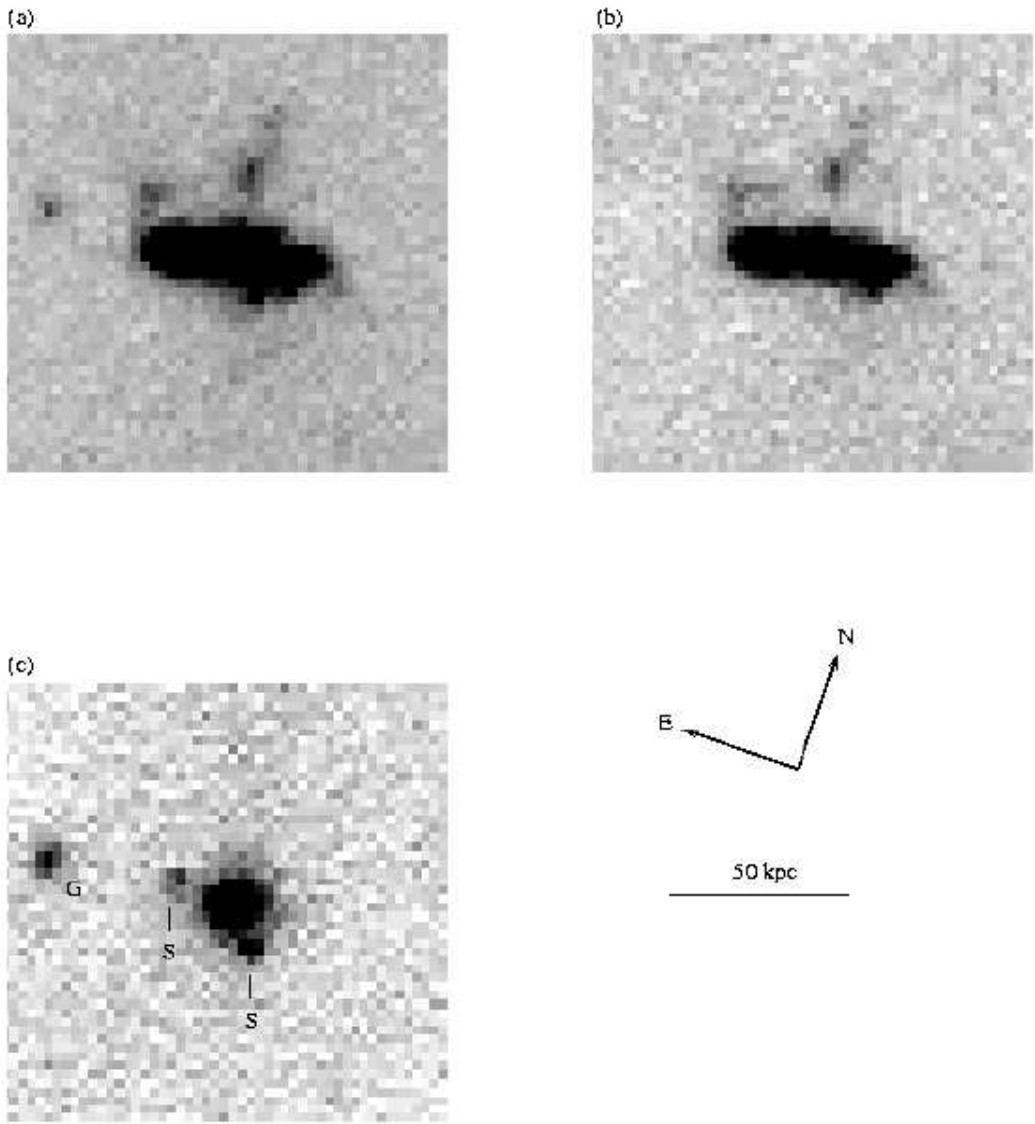,width=15cm}
\caption{TTF images of 3C171: (a) H$\alpha+$continuum; (b) pure  H$\alpha$; (c) continuum ($\lambda_c =$8570\AA\,). The symbols have the same meaning as in 
Figure 1.
}
\end{figure}
\begin{figure}
\psfig{figure=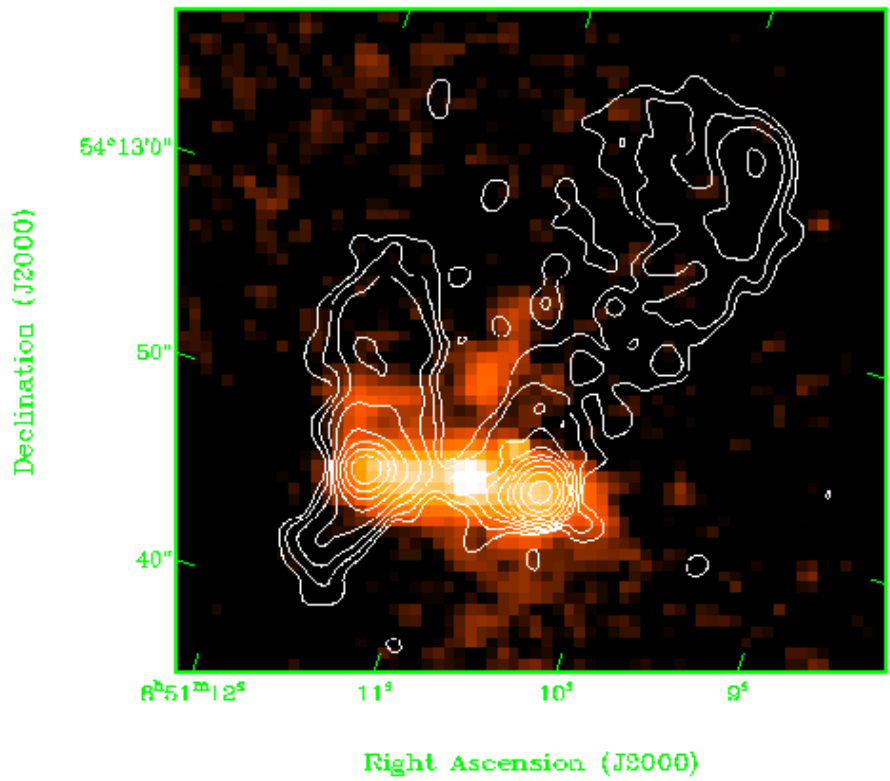,width=15cm}
\caption{Overlay of the H$\alpha$ image for 3C171 (greyscale) 
with the
6cm radio map of Blundell(1996) (contours).
}
\end{figure}

\end{document}